\documentclass[aps, prc,reprint, superscriptaddress,linenumbers]{revtex4}
\usepackage{amsfonts}
\usepackage{amssymb}
\usepackage{float}
\usepackage{amsmath}
\usepackage[pdftex]{hyperref}
\usepackage{graphicx,color}
\usepackage{dcolumn}
\usepackage{bm}
\usepackage{here}
\usepackage{comment}
\usepackage[paperwidth=210mm,paperheight=297mm,centering,hmargin=1.5cm,vmargin=2.0cm]{geometry}
\usepackage{tabularx}
\usepackage{caption}
\usepackage{subcaption}
\usepackage{multirow}

\allowdisplaybreaks[1]

\makeatletter
\newsavebox{\@brx}
\newcommand{\llangle}[1][]{\savebox{\@brx}{\(\m@th{#1\langle}\)}%
  \mathopen{\copy\@brx\mkern2mu\kern-0.9\wd\@brx\usebox{\@brx}}}
\newcommand{\rrangle}[1][]{\savebox{\@brx}{\(\m@th{#1\rangle}\)}%
  \mathclose{\copy\@brx\mkern2mu\kern-0.9\wd\@brx\usebox{\@brx}}}
\makeatother

\newcommand{\muB}{$\mu_{\rm B}$}
\newcommand{\sNN}{$\sqrt {{s_{\rm NN}}}~$}

\begin{document}

\title{Efficiency corrections for mutually inclusive variables and particle identification effect for mixed-cumulants in heavy-ion collisions} 

\author{Arghya Chatterjee}
\email{arghya@mail.ccnu.edu.cn}
\affiliation{Key\,Laboratory\,of\,Quark\,\&\,Lepton\,Physics\,(MOE)\,and\,Institute\,of\,Particle\,Physics,\,Central\,China\,Normal\,University,\,Wuhan\,430079,\,China}
\author{Toshihiro Nonaka}
\email{nonaka.toshihiro.ge@u.tsukuba.ac.jp}
\affiliation{Tomonaga\,Center\,for\,the\,History\,of\,the\,Universe,\,University\,of\,Tsukuba,\,Tsukuba,\,Ibaraki\,305,\,Japan}
\author{ShinIchi Esumi}
\email{esumi.shinichi.gn@u.tsukuba.ac.jp}
\affiliation{Tomonaga\,Center\,for\,the\,History\,of\,the\,Universe,\,University\,of\,Tsukuba,\,Tsukuba,\,Ibaraki\,305,\,Japan}
\author{Xiaofeng Luo}
\email{xfluo@ccnu.edu.cn}
\affiliation{Key\,Laboratory\,of\,Quark\,\&\,Lepton\,Physics\,(MOE)\,and\,Institute\,of\,Particle\,Physics,\,Central\,China\,Normal\,University,\,Wuhan\,430079,\,China}


\begin{abstract}
Mix-cumulants of conserved charge distributions are sensitive observables for probing properties of the QCD medium and phase transition in heavy-ion collisions. To perform precise measurements, efficiency correction is one of the most important step. In this study, using the binomial efficiency model, we derive efficiency correction formulas for mutually exclusive and inclusive variables. The UrQMD model is applied to verify the validity of these formulas for different types of correlations. Furthermore, we investigate the effect of the multiplicity loss and contamination emerging from the particle identifications. This study provides important step toward future measurements of mixed-cumulants in relativistic heavy-ion collisions. 
\end{abstract}
\maketitle

\newcommand{\ave}[1]{\ensuremath{\langle#1\rangle} }
\newcommand{\avebig}[1]{\ensuremath{\Bigl\langle#1\Bigr\rangle} }
\newcommand{\aveave}[1]{\ensuremath{\langle\!\langle#1\rangle\!\rangle} }

\onecolumngrid

\section{Introduction}
Heavy-ion collisions at relativistic energies produce matter at extreme energy density and temperature conditions. 
This matter is likely to comprise deconfined quarks and gluons and is called quark-gluon plasma (QGP). 
A primary objective of heavy-ion collision experiments is to explore the phase structure of the hot dense QCD matter. 
The QCD phase structure can be expressed as a function of temperature ($T$) and baryon chemical potential (\muB)~\cite{Rajagopal:1999cp}.
QCD based model calculations predict that at large \muB~values the transition from hadronic matter to QGP is of the first order~\cite{Stephanov:2007fk,Bowman:2008kc}. 
The end point of the first order phase transition boundary is known as the QCD critical point (CP), after which there is no genuine phase transition, except a smooth crossover from hadronic to quark-gluon degrees of freedom~\cite{Aoki:2006we,Gupta:2011wh}. 
One of the major approaches to exploring the QCD phase structure is via the measurements of event-by-event fluctuations of conserved quantities, such as net-baryon ($B$), net-charge ($Q$) and net-strangeness ($S$)~\cite{Stephanov:1998dy,Bzdak:2019pkr,Luo:2017faz,Chatterjee:2012np,Garg:2013ata}. 
In a thermodynamic system, $r$-th order fluctuations (''cumulants") of event-by-event net-multiplicity distributions are related to the $r$-th order thermodynamic susceptibilities of the corresponding conserved charges that diverge near the critical point~\cite{Stephanov:2008qz, Cheng:2008zh, Asakawa:2009aj,Stephanov:2011pb}. 
Furthermore, these measured cumulants have also been used to extract freeze-out parameters ($T$ and $\mu_{\text{B}}$) by comparing them with model calculations from lattice QCD and hadron resonance gas (HRG)~\cite{Bazavov:2012vg,Borsanyi:2013hza,Alba:2014eba,Flor:2020fdw,Gupta:2020pjd}. 
Owing to the experimental constraints on measuring neutral particle yields, net-proton and net-kaons are adopted as experimental proxies of net-baryon and net-strangeness respectively. In the last decade, the STAR and PHENIX experiments at the relativistic heavy ion collider (RHIC) have measured the second, third and forth order cumulants of net-proton~\cite{Aggarwal:2010wy,Luo:2015doi, Adamczyk:2013dal, Adam:2020unf,Abdallah:2021fzj}, net-charge~\cite{Adamczyk:2014fia, Adare:2015aqk} and net-kaon~\cite{Adamczyk:2017wsl} multiplicity distributions over a wide range of collision energies to determine non-monotonic energy dependence behaviours, as an indicator of the CP presence. Within current statistical uncertainties, no distinctive signatures of the CP have been inferred from the net-charge and net-kaon measurements. However, recent measurements of forth order to second order cumulant ratios of net-proton multiplicity distributions exhibit non-monotonic energy dependence as a function of \sNN~\cite{Adam:2020unf}. Although, before drawing any substantial physics conclusion from event-by-event fluctuation measurements, we need to carefully investigate the different background contributions~\cite{Luo:2013bmi, Chatterjee:2019fey, Zhang:2019lqz, Westfall:2014fwa, Zhou:2018fxx,Chatterjee:2020nnn}. 

Recently, the STAR experiment reported the first measurement of second-order mixed-cumulants between net-charge, net-proton, and net-kaon multiplicity distributions in the first phase of the beam energy scan (BES-I) program at RHIC~\cite{Adam:2019xmk}. 
These mixed-cumulants are related to the off-diagonal thermodynamic susceptibilities that carry the correlation between different conserved charges of QCD~\cite{Koch:2005vg,Gavai:2005yk,Majumder:2006nq,Bluhm:2008sc, Ding:2015fca,Chatterjee:2016mve,Bellwied:2019pxh}. 
The importance of the second-order mixed-cumulants was first highlighted in the context of normalized baryon-strangeness susceptibilities ($C_{B,S} = -3\chi^{1,1}_{B,S}/\chi^2_{S}$) in Ref.~\cite{Koch:2005vg}, which are expected to exhibit a rapid change with the onset of deconfinement. 
These quantities can be investigated by measuring the energy dependence ratios of off-diagonal over diagonal cumulant ratios between net-baryon and net-strangeness ($C^{1,1}_{B,S}/C^{2}_{S}$). 
Another research objective originates from the comparisons between the ideal HRG model and lattice QCD calculations.
The baryon-charge susceptibility ($\chi^{1,1}_{B,Q}$) exhibits a significant difference between the lattice and ideal HRG calculations above the crossover transition temperature, even in at the lowest order~\cite{Bazavov:2012jq,Vovchenko:2016rkn}. A similar difference between the lattice and HRG calculations can also be observed in higher-order baryon susceptibilities (($\chi^{4}_B$)), which is more statistically challenged in the experimental measurement~\cite{Karsch:2017zzw}. 
Similar to diagonal cumulants, the mixed-cumulants are also limited by the lack of neutral particle detection capability. 
The measurements of charge-baryon or charge-strangeness mixed-cumulants are less affected by such an experimental limitation, as the neutral particles do not contribute to such charge correlations, and can be approximated by $C^{1,1}_{Q,B} \approx C^{1,1}_{Q,p}$ and $C^{1,1}_{Q,S}\approx C^{1,1}_{Q,k}$~\cite{Chatterjee:2016mve}. 
In contrast, baryon-strangeness mixed-cumulants cannot be approximated by the proton-kaon off-diagonal. However, the relationship between these cumulants have been studied in Ref.~\cite{Chatterjee:2016mve,Yang:2016xga}. 

Recent measurements of second-order mixed-cumulants at the RHIC energy range (\sNN = 7.7-200 GeV) agree well with different model predictions for the net proton-kaon off-diagonal correlator ($C_{p,k} = \sigma^{1,1}_{p,k}/\sigma^{2}_{k}$). 
However, the charge-proton ($C_{Q,p} = \sigma^{1,1}_{Q,p}/\sigma^{2}_{p}$) and charge-kaon ($C_{Q,k} = \sigma^{1,1}_{Q,k}/\sigma^{2}_{k}$) correlators significantly deviate from the model predicted values. In this study, we demonstrate that the deviations observed in charge-proton and charge-kaon correlators are owing to the efficiency double counting.
We argue that, to correct the effects of efficiency for the charge-proton and charge-kaon mixed-cumulants, the particle identification for charge needs to be performed with the estimation of corresponding efficiencies. 
However, a significant number of charged tracks are missed for particle identification as different detectors used~\cite{Adam:2019xmk}. 
In this study, we focus on the 2nd-order mixed-cumulant for two variables to elucidate and simplify 
several important points on the efficiency correction. 
Although the points also apply for higher-order mixed-cumulants, including more than two variables cases, 
these extensions should be straightforward and are expected to be investigated in future studies. 

This paper is organized as follows. 
In Sec~\ref{sec:effcorr}, cumulants, mixed-cumulants, and their efficiency 
corrections are introduced. 
The formulas for the efficiency correction of the 2nd-order mixed-cumulant is discussed 
for two types of correlations.
In Sec.~\ref{sec:UrQMD}, we perform numerical analysis using the UrQMD model 
to verify the importance of adopting appropriate formulas, 
depending on the correlation type.
Here, the effects of double-counting are discussed, 
and the potential effects of the multiplicity loss owing to particle identification are investigated.
Finally, we summarize this study in Sec.~\ref{sec:Summary}.

\section{Efficiency correction\label{sec:effcorr}}
\subsection{Cumulants and mixed-cumulants}
In statistics, any distribution can be characterized by different order moments or cumulants. 
The $r$th-order moment of variable $N$ is defined by the $r$th order derivative of moment generating function 
$G(\theta)$:
\begin{eqnarray}
	G(\theta) &=& \sum_{N}e^{N\theta}P(N) = \ave{e^{N\theta}}, \label{eq:G}\\
	\ave{N^{r}} &=& \frac{d^{r}}{d\theta^{r}}G(\theta)\Bigl|_{\theta=0}, \label{eq:mu}
\end{eqnarray}
where $P(N)$ is a probability distribution function, 
and $\ave{\cdot}$ represents an average over events. 
Cumulants are defined by the cumulant generating function $T(\theta)$, which is the logarithm of moment generating function:
\begin{eqnarray}
	T(\theta) &=& {\rm ln}G(\theta), \label{eq:K}\\
	\ave{N^{r}}_{\rm c} &=& \frac{d^{r}}{d\theta^{r}}T(\theta)\Bigl|_{\theta=0}, \label{eq:kappa} 
\end{eqnarray}
where $\ave{\cdot}_{\rm c}$ represents the cumulant of the variable inside the bracket.
From Eqs.~\ref{eq:G}--\ref{eq:kappa}, the 1st and 2nd-order cumulants are expressed in terms of moments 
\begin{eqnarray}
	\ave{N}_{\rm c}&=& \ave{N}, \\
	\ave{N^{2}}_{\rm c}&=& \ave{N^{2}} - \ave{N}^{2},
\end{eqnarray}
Similarly, the multivariate moments and cumulants are defined by the multivariate generating function.    
In this study, we focus on the two-variable case, which we call "mixed-" moments or cumualnts, where the moment generating function is given by 
\begin{eqnarray}
	G(\theta_{1},\theta_{2}) 
	&=& \sum_{N_{1},N_{2}}e^{\theta_{1}N_{1}}e^{\theta_{2}N_{2}}P(N_{1},N_{2})
	= \ave{e^{\theta_{1}N_{1}}e^{\theta_{2}N_{2}}}, \label{eq:Gmix}\\
	\ave{N_{1}^{r_{1}}N_{2}^{r_{2}}} &=&  
	\frac{\partial^{r_{1}}}{\partial\theta_{1}^{r_{1}}}\frac{\partial^{r_{2}}}{\partial\theta_{2}^{r_{2}}}
	G(\theta_{1},\theta_{2})\Big|_{\theta_{1}=\theta_{2}=0}. \label{eq:mumix}
\end{eqnarray}
Mixed-cumulants are then defined as
\begin{eqnarray}
	T(\theta_{1},\theta_{2}) &=& {\rm ln}G(\theta_{1},\theta_{2}), \label{eq:Kmix}\\
	\ave{N_{1}^{r_{1}}N_{2}^{r_{2}}}_{\rm c} &=&
	\frac{\partial^{r_{1}}}{\partial\theta_{1}^{r_{1}}}\frac{\partial^{r_{2}}}{\partial\theta_{2}^{r_{2}}}
	T(\theta_{1},\theta_{2})\Big|_{\theta_{1}=\theta_{2}=0}. \label{eq:kappamix} 
\end{eqnarray}
From Eqs.~\ref{eq:Gmix}--\ref{eq:kappamix}, we obtain the 2nd-order 
mixed-cumulant in terms of moments and mixed-moments: 
\begin{eqnarray}
	\ave{N_{1}N_{2}}_{\rm c} = \ave{N_{1}N_{2}} - \ave{N_{1}}\ave{N_{2}}.
\end{eqnarray}

\subsection{Binomial model}
The particle detection efficiency of each detector is always limited.
The event-by-event particle multiplicity distributions are convoluted owing to this finite detector efficiency. 
The efficiency correction needs to be performed to recover the true multiplicity distribution. 
For simplicity, we assume that the detection
efficiency can be approximated by the binomial efficiency response function~\cite{Bzdak:2012ab,Bzdak:2013pha}.
The mean value (first-order moment/cumulant) can be easily reconstructed by division with the binomial efficiency response; however, its influence on higher-order cumulants is complicated and depends on the probability distribution of efficiency~\cite{binomial_breaking,Nonaka:2018mgw,Esumi:2020xdo}. 
Throughout this paper, we focus on a simple assumption of the binomial distribution given by 
\begin{eqnarray}
	\tilde{P}(n) &=& \sum_{N}P(N)B_{\varepsilon,N}(n), \\ 
	B_{\varepsilon,N}(n) &=& \frac{N!}{n!(N-n)!}\varepsilon^{n}(1-\varepsilon)^{N-n},
        \label{eq:P=PB}
\end{eqnarray}  
where $\varepsilon$ represents the efficiency, while $N$ and $n$ are generated and measured particles, respectively.
In this case, the correction formulas can be derived in a straightforward manner as discussed in the literature
~\cite{Bialas:1985jb,eff_kitazawa,eff_koch,eff_psd_volker,eff_xiaofeng,eff_psd_kitazawa,Nonaka:2017kko,Luo:2018ofd}. 
The efficiency correction for the 2nd-order mixed-cumulant is given by~\cite{Nonaka:2017kko,Luo:2018ofd}
\begin{eqnarray}
	  \aveave{ K_{(x)}K_{(y)} }_{\rm c}
  &=& \ave{\kappa_{(1,0,1)}\kappa_{(0,1,1)}}_{\rm c} + \ave{\kappa_{(1,1,1)}}_{\rm c} - \ave{\kappa_{(1,1,2)}}_{\rm c},
	\label{eq:QQ11}
\end{eqnarray}
with 
\begin{eqnarray}
	K_{(x)} = \sum_{i}^{M}x_{i}n_{i},\;&& K_{(y)} = \sum_{i}^{M}y_{i}n_{i}, \label{eq:multi_Q}\\
	\kappa{(r,s,t)} &=& 
	\sum_{i=1}^{M}\frac{x_{i}^{r}y_{j}^{s}}{\varepsilon_{i}^{t}}n_{i}. 
        \label{eq:multi_qmix}
\end{eqnarray}
where $\aveave{\cdot}$ represents the efficiency correction, 
$M$ is the number of efficiency bins, 
$n_{i}$ indicates the number of particles, $\varepsilon_{i}$ is the efficiency, 
$x_{i}$ and $y_{i}$ are the electric charge of the particles at the $i$th efficiency bin.
Notably Eqs.~\ref{eq:multi_Q} and \ref{eq:multi_qmix} can be 
rewritten in track-by-track notations as
\begin{eqnarray}
	K_{(x)} = \sum_{j}^{n^{\rm tot}}x_{j},\;&& K_{(y)} = \sum_{i}^{n^{\rm tot}}y_{j}, \label{eq:multi_Q_tbt} \\
	\kappa{(r,s,t)} &=& 
	\sum_{j=1}^{n^{\rm tot}}\frac{x_{j}^{r}y_{j}^{s}}{\varepsilon_{j}^{t}}, 
        \label{eq:multi_qmix_tbt}
\end{eqnarray}
where $n_{\rm tot}=\sum_{i}^{M}n_{i}$ is the number of measured particles in one event, 
and the other variables are now track-wise with $j$ running over the particles in the summation. 

In the rest of this section, we consider two efficiency bins for simplicity. 
Particles for each bin have the same efficiency values, $\varepsilon_{1}$ and $\varepsilon_{2}$, respectively. 
The number of particles will be denoted by $n_{1}$ and $n_{2}$.

\subsection{Mutually exclusive variable}
Let's consider the correlation between two mutually exclusive variables.  
From Eq.~\ref{eq:kappamix}, the mixed-cumulant of generated particles are expanded in terms of moments as 
\begin{equation}
	\ave{N_{1}N_{2}}_{\rm c} = \ave{N_{1}N_{2}} - \ave{N_{1}}\ave{N_{2}}. \label{eq:mix2_true}
\end{equation}
To perform the efficiency correction relative to measured particles, we suppose
\begin{eqnarray}
	x &=& (x_{1},x_{2}) = (1,0), \\	
	y &=& (y_{1},y_{2}) = (0,1), 	
\end{eqnarray}
to consider $\aveave{K_{(x)}K_{(y)}}_{\rm c}$ with $K_{(x)}=n_{1}$ and $K_{(y)}=n_{2}$. 
From Eq.~\ref{eq:QQ11}, we get
\begin{eqnarray}
  \aveave{ K_{(1,0)}K_{(0,1)} }_{\rm c}
  &=& \ave{\kappa{(1,0,1)}\kappa{(0,1,1)}}_{\rm c} + \ave{\kappa{(1,1,1)}}_{\rm c} - \ave{\kappa{(1,1,2)}}_{\rm c} \\
  &=& \avebig{\frac{n_{1}}{\varepsilon_{1}}\frac{n_{2}}{\varepsilon_{2}}}_{\rm c} \\
  &=& \frac{1}{\varepsilon_{1}\varepsilon_{2}}\ave{n_{1}n_{2}} - \frac{1}{\varepsilon_{1}\varepsilon_{2}}\ave{n_{1}}\ave{n_{2}}, \label{eq:mix2_corr}
\end{eqnarray}
which is the basic formula of the efficiency correction 
for the 2nd-order mixed-cumulant of two variables.

\subsection{Mutually inclusive variables} 
\subsubsection{Problem}
Next, we consider the correlation between $N_{1}$ and $N_{1}+N_{2}$, i.e, when the two variables are not mutually exclusive. 
It is clear that we have the self-correlation of $N_{1}$. 
The 2nd-order mixed-cumulant can be expanded as 
\begin{eqnarray}
	\ave{N_{1}(N_{1}+N_{2})}_{\rm c} 
	&=& \ave{N_{1}N_{2}} - \ave{N_{1}}\ave{N_{2}} + \ave{N_{1}^{2}} - \ave{N_{1}}^{2}, \label{eq:mix2_self_true}
\end{eqnarray}
where the last two terms represent the variance (2nd-order cumulant, $\ave{N_{1}^{2}}_{\rm c}$).
If we employ Eq.~\ref{eq:mix2_corr} for the efficiency correction, 
we can just replace $n_{1}$ to $n_{1}+n_{2}$ as
\begin{eqnarray}
	\aveave{K_{(1,0)}K_{(0,1)}}_{\rm c} 
	&=& \frac{1}{\varepsilon_{1}\varepsilon_{2}'}\ave{n_{1}(n_{1}+n_{2})} - \frac{1}{\varepsilon_{1}\varepsilon_{2}'}\ave{n_{1}}\ave{n_{1}+n_{2}}, \\
	&=& \frac{1}{\varepsilon_{1}\varepsilon_{2}'}\ave{n_{1}n_{2}} - \frac{1}{\varepsilon_{1}\varepsilon_{2}'}\ave{n_{1}}\ave{n_{2}}
	+ \frac{1}{\varepsilon_{1}\varepsilon_{2}'}\ave{n_{1}^{2}} - \frac{1}{\varepsilon_{1}\varepsilon_{2}'}\ave{n_{1}}^{2}, \label{eq:mix2_self_wrong}
\end{eqnarray}
with $\varepsilon_{2}'$ being the averaged efficiency for $N_{1}$ and $N_{2}$ given by
\begin{eqnarray}
	\varepsilon_{2}' = \frac{\ave{N_{1}}\varepsilon_{1}+\ave{N_{2}}\varepsilon_{2}}{\ave{N_{1}}+\ave{N_{2}}},
\end{eqnarray}
which is not an appropriate efficiency corrected expression for mutually inclusive variables. 
To confirm this, we suppose that two independent variables $N_{1}$ and $N_{2}$ follow the Poisson distribution 
with $\ave{N_{1}}=4$ and $\ave{N_{2}}=6$. 
It is known that the relation $\ave{N^{r}}_{\rm c} = \ave{N}_{\rm c}$ holds for Poisson distributions; 
thus, $\ave{N^{2}}_{\rm c}=\ave{N^{2}}-\ave{N}^{2}$, which leads to $\ave{N_{1}(N_{1}+N_{2})}_{\rm c}=4$ ,
from Eq.~\ref{eq:mix2_self_true}. 
The relationship $\ave{N_{1}N_{2}}=\ave{N_{1}}\ave{N_{2}}$ was adopted for the independent variables.
We then consider the efficiency correction for $\varepsilon_{1}=0.5$ and $\varepsilon_{2}=0.4$. 
Accordingly,
\begin{eqnarray}
	&& \varepsilon_{2}' = 0.44,\; \ave{n_{1}}=2,\; \ave{n_{2}}=2.4, \\
	&& \ave{n_{1}^{2}}_{\rm c} = \ave{n_{1}^{2}} - \ave{n_{1}}^{2} = 2, \\
	&& \ave{n_{1}n_{2}} = \ave{n_{1}}\ave{n_{2}}, 
\end{eqnarray}
which leads to the efficiency corrected mixed-cumulant value 
\begin{eqnarray}
	\aveave{K_{(1,0)}K_{(0,1)}}_{\rm c} = 4.55. 
\end{eqnarray}
Hence, Eq.~\ref{eq:mix2_corr} is not valid for the self correlated or mutually inclusive case. 

\subsubsection{Solution}
The solution is to adopt the appropriate indices for $x$ and $y$ in Eqs.~\ref{eq:multi_qmix} and \ref{eq:multi_Q}.
To consider $\aveave{K_{(x)}K_{(y)}}_{\rm c}$ with $K_{(x)}=n_{1}$ and $K_{(y)}=n_{1}+n_{2}$, 
the indices should have been
\begin{eqnarray}
	x &=& (x_{1},x_{2}) = (1,0), \\	
	y &=& (y_{1},y_{2}) = (1,1), 	
\end{eqnarray}
thus
\begin{eqnarray}
  \aveave{ K_{(1,0)}K_{(1,1)} }_{\rm c}
  &=& \ave{\kappa{(1,0,1)}\kappa{(0,1,1)}}_{\rm c} + \ave{\kappa{(1,1,1)}}_{\rm c} - \ave{\kappa{(1,1,2)}}_{\rm c} \\
  &=& \frac{1}{\varepsilon_{1}\varepsilon_{2}}\ave{n_{1}n_{2}} - \frac{1}{\varepsilon_{1}\varepsilon_{2}}\ave{n_{1}}\ave{n_{2}}
	+ \frac{1}{\varepsilon_{1}^{2}}\ave{n_{1}^{2}} - \frac{1}{\varepsilon_{1}^{2}}\ave{n_{1}}^{2}
	+ \frac{1}{\varepsilon_{1}}\ave{n_{1}} - \frac{1}{\varepsilon_{1}^{2}}\ave{n_{1}}, \label{eq:mix2_self_correct}
\end{eqnarray}
where we determine two additional terms compared with Eq.~\ref{eq:mix2_self_wrong}. 
It is inferred that the last four terms in Eq.~\ref{eq:mix2_self_correct} represent 
the efficiency correction of the variance (2nd-order cumulant), $\aveave{K_{(x)}^{2}}$, 
which corresponds to the last two terms in Eq.~\ref{eq:mix2_self_true}
\footnote{This can be confirmed by substituting $(x_{1},x_{2})=(1,0)$ and $(y_{1},y_{2})=(1,0)$ into Eqs.~\ref{eq:QQ11}--\ref{eq:multi_qmix}}.
This indicates that the variance has to be correctly considered for the mutually inclusive variable case, 
which cannot be handled by Eq.~\ref{eq:mix2_corr}.

We summarize this section as follows. 
The efficiency correction formula for the 2nd-order mixed-cumulant was fully expanded for two cases: 
one is for two mutually exclusive variables, 
and the other case assumes that one variable is a subset of the other, 
to consider the self-correlation, 
as expressed in Eqs.~\ref{eq:mix2_corr} and \ref{eq:mix2_self_correct}. 
Both cases were determined to be incompatible with each other. 
The proper correction formulas needs to be obtained
by substituting appropriate indices into Eqs.~\ref{eq:QQ11}--\ref{eq:multi_qmix}.
This implies that the efficiency values have to be handled properly  
for each variable, without averaging them, especially when considering the self-correlation. 
It should be noted that the risk of using the averaged efficiency has already been pointed out in Ref.~\cite{Nonaka:2017kko} for higher-order cumulants of single-variables. 
The efficiency bins always need to be carefully handled. 
The track-by-track efficiency via the identified particle approach expressed in Eqs.~\ref{eq:multi_Q_tbt} and \ref{eq:multi_qmix_tbt} would be a better way to handle all possible variations of efficiencies~\cite{Luo:2018ofd}. 
However, the particle identification needs to be applied to determine the efficiencies for different particle species, which does not discard a small amount of particles, depending on the overlapping area of the variables for the particle identification. This effect will be studied by numerical simulations in the next section~\cite{Shao:2005iu,Adamczyk:2014fia}.

\section{Numerical analysis in UrQMD model\label{sec:UrQMD}}



\subsection{Closure test using UrQMD model}

\begin{figure*}[htp!]
	\centering 
	\includegraphics[width=0.95\textwidth]{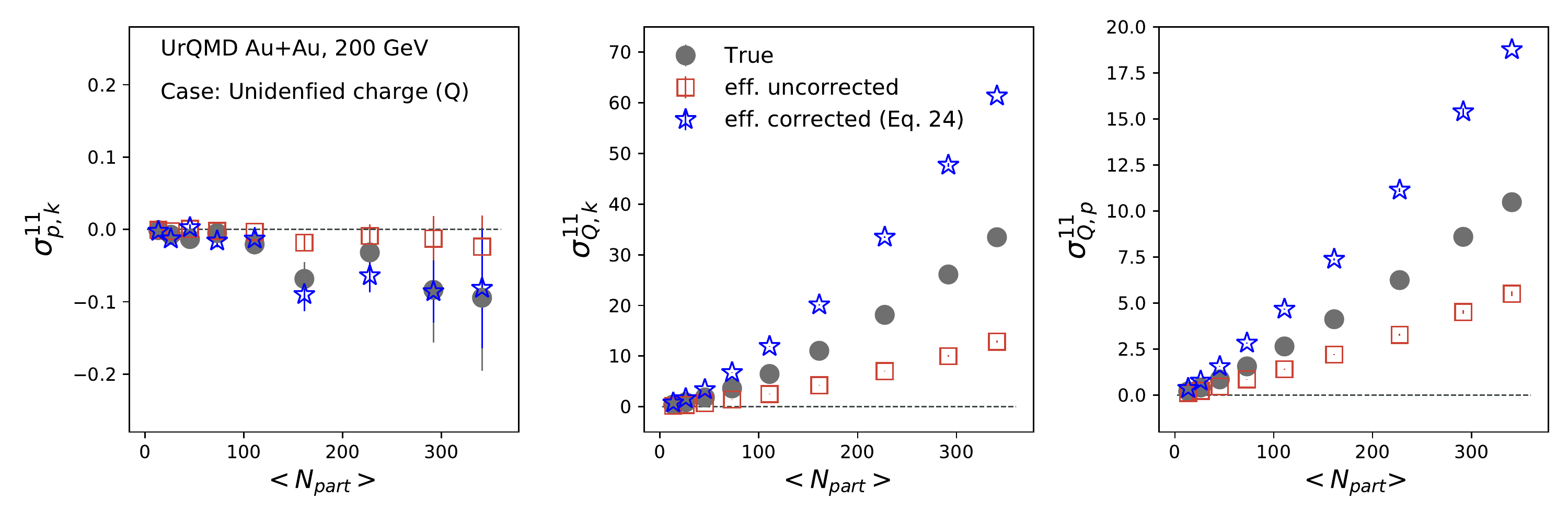}
	\caption{(Color online) Centrality dependence of second-order mixed-cumulants of net-charge ($Q=N_{Q^{+}}-N_{Q^{-}}$), net-proton ($p=N_{p}-N_{\bar{p}}$) and net-kaon ($k=N_{k^{+}}-N_{k^{-}}$) multiplicities for Au+Au collisions at 200 GeV, using the UrQMD model. The efficiency corrections are performed assuming variables are mutually exclusive (Eq. \ref{eq:mix2_corr}).}
	\label{offdiaOld}
\end{figure*}

To validate the discussion from the previous section, we have analyzed the second-order mixed cumulants from the UrQMD event generator at \sNN =  200 GeV. 
The UrQMD is a microscopic transport model, where the space-time evolution of the fireball is considered in terms of the excitation of color strings that fragment further into hadrons, the co-variant propagation of hadrons and resonances that undergo scatterings, and finally the decay of all the resonances~\cite{Bass:1998ca, Bleicher:1999xi}. 
The collision energy dependence of the baryon stopping phenomenon is dynamically incorporated in the UrQMD model. 
The UrQMD model has been relatively successful and widely used in the heavy-ion phenomenology~\cite{Bleicher:1999xi, Bleicher:1998wu}. 
Previously, this model was adopted to study several observable fluctuation and cumulants~\cite{Bleicher:2000ek, Haussler:2005ei, Luo:2013bmi, Chatterjee:2016mve, Xu:2016qjd, He:2017zpg, Mukherjee:2017elm}. 
More information on the UrQMD model can be found in Ref.~\cite{Bass:1998ca, Bleicher:1999xi}. 
In this study, we have adopted approximately one million events for Au+Au collisions at \sNN = 200 GeV to probe the efficiency correction effect on mixed cumulants. 
The obtained results are presented for 9 different centrality bins represented by the average number of participant nucleons ($\ave{N_{part}}$). 
In this study, we have applied the same kinematic acceptance $|\eta|<0.5$ and $0.4<p_{T}<1.6$ GeV/$c$ as STAR data~\cite{Adam:2019xmk}. 
The collision centrality is defined using RefMult2 (charged particle multiplicity within the pseudorapidity range $0.5<|\eta|<1.0$) to reduce the centrality auto-correlation effect~\cite{Luo:2013bmi,Chatterjee:2019fey}. 
Figure~\ref{offdiaOld} illustrates the centrality dependence of second-order mixed-cumulants of net-charge, net-proton, and net-kaon multiplicity distributions from the UrQMD model. 
The gray solid points represent the 'true' mixed-cumulants values. 
To introduce the detector efficiency effect, we passed the binomial filter to the counted particles number in each event. 
We adopted two $p_{T}$-bin and positive-negative separate efficiencies similar to real data analysis. We took this approach because different detector subsystems are used for particle identification in a high or low momentum region~\cite{Adam:2019xmk}.  
These subsystems have different efficiencies, and it is always better to use their proper efficiency values over their average efficiencies~\cite{Nonaka:2017kko}. 
Subsequently, we estimated the mixed-cumulants with filtered particle numbers represented by red square points, which are analogous to efficiency uncorrected values.

\begin{figure*}[htp!]
	\centering 
	\includegraphics[width=0.95\textwidth]{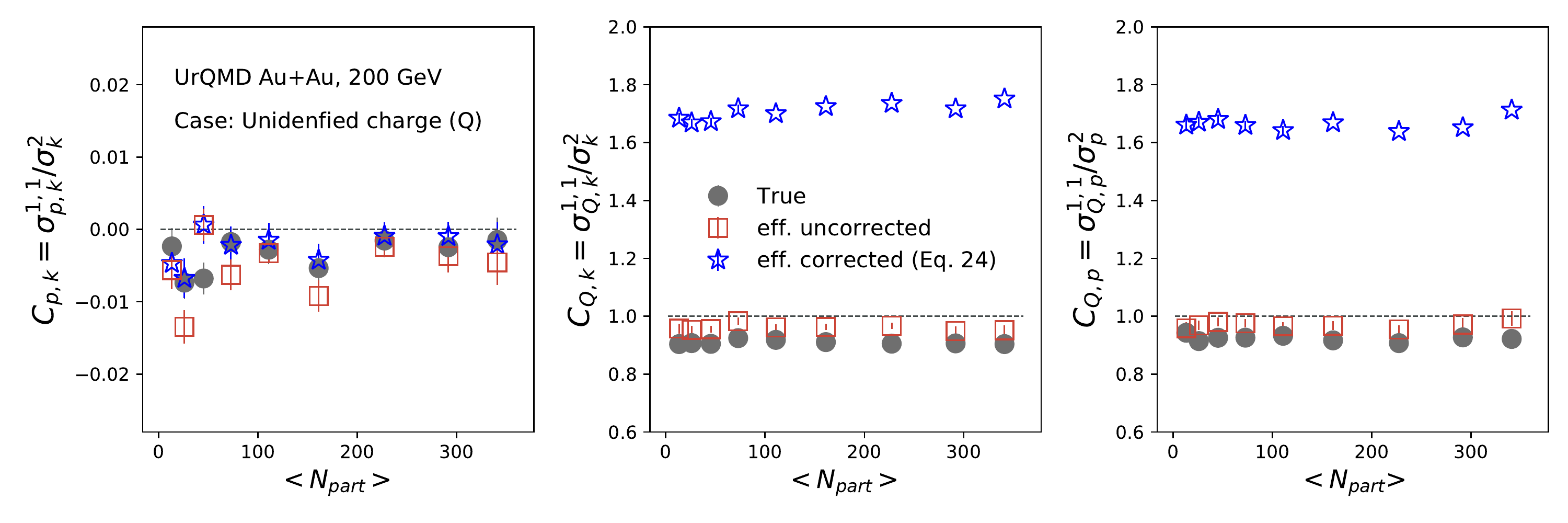}
	\caption{(Color online) Centrality dependence of second-order off-diagonal to diagonal cumulant ratios for Au+Au collisions at 200 GeV, using the UrQMD model. The efficiency correction are performed assuming the variables are mutually exclusive (Eq. \ref{eq:mix2_corr}).}
	\label{ratioOld}
\end{figure*}

\begin{figure*}[htp!]
	\centering 
	\includegraphics[width=0.95\textwidth]{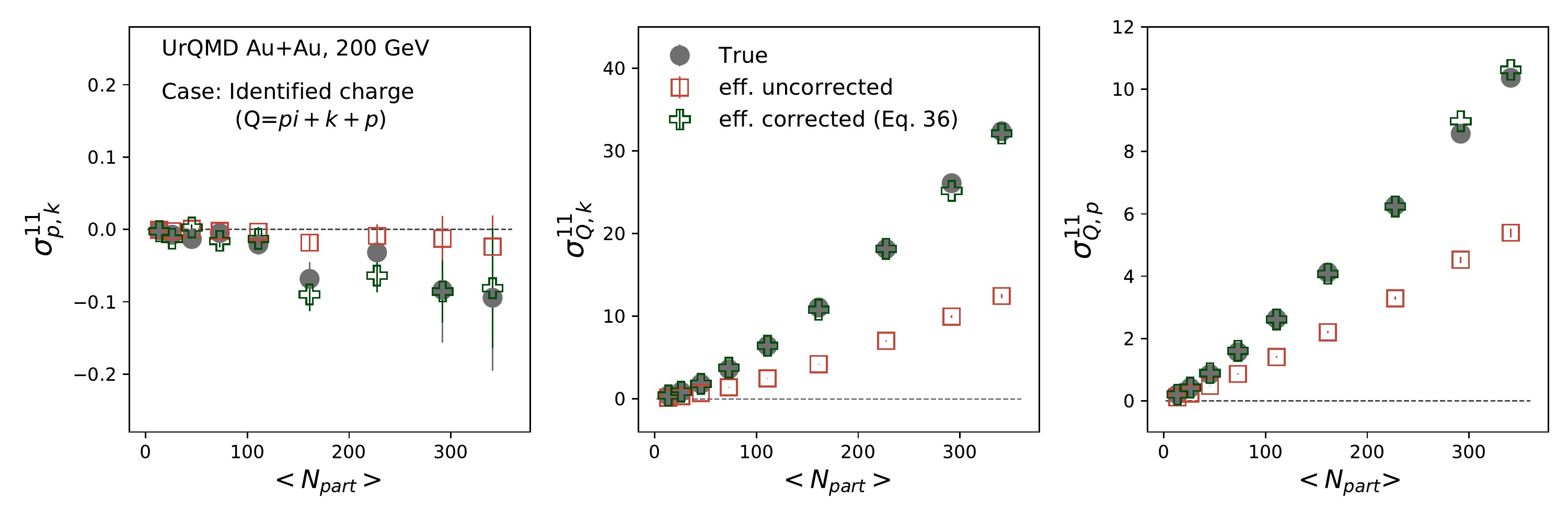}
	\caption{(Color online) Centrality dependence of second-order mixed-cumulants of identified net-charge ($Q=(N_{\pi^{+}}+N_{k^{+}}+N_{p})-(N_{\pi^{-}}+N_{k^{-}}+N_{\bar{p}})$), net-proton ($p=N_{p}-N_{\bar{p}}$), and net-kaon ($k=N_{k^{+}}-N_{k^{-}}$) multiplicity for Au+Au collisions at 200 GeV, using the UrQMD model. The efficiency corrections are done assuming the variables are mutually inclusive (Eq. \ref{eq:mix2_self_correct}).}
	\label{offdiaNew}
\end{figure*}

\begin{figure*}[htp!]
	\centering 
	\includegraphics[width=0.95\textwidth]{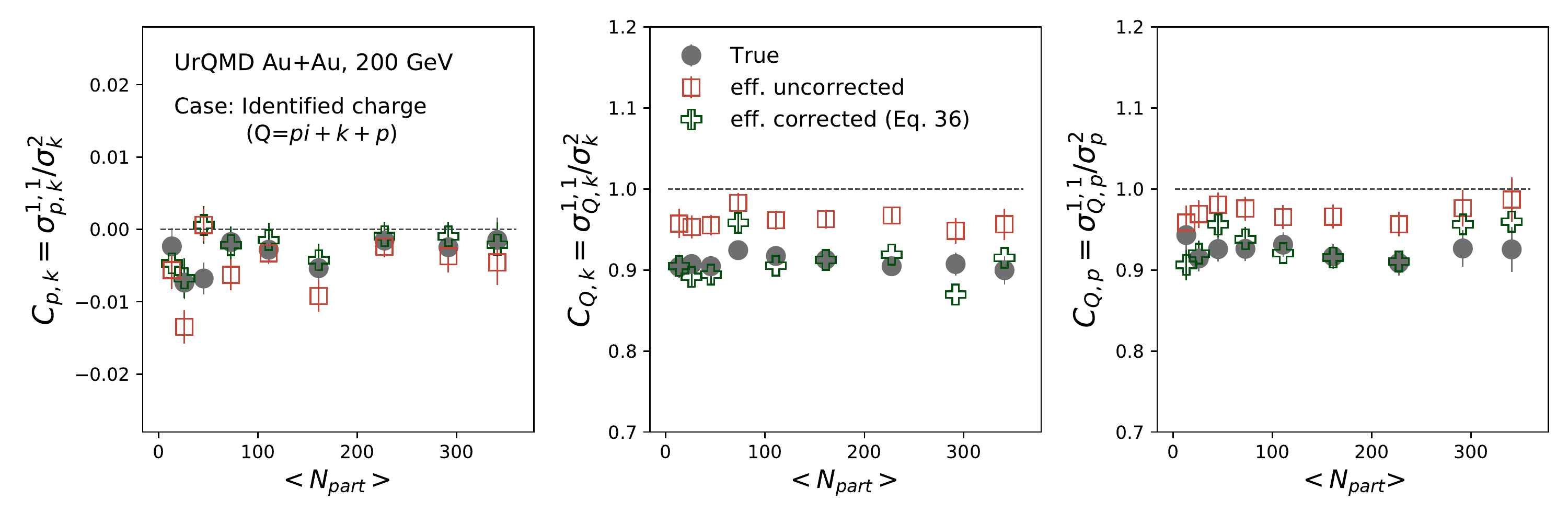}
	\caption{(Color online) Centrality dependence of second-order off-diagonal to diagonal cumulant ratios of identified charged particles for Au+Au collisions at 200 GeV, using the UrQMD model. The efficiency correction are performed assuming the variables are mutually inclusive (Eq. \ref{eq:mix2_self_correct}).}
	\label{ratioNew}
\end{figure*}

In the next step, we correct the efficiency using input efficiency values.
In this case, we adopted unidentified charged particles for the net-charge ($Q$) and applied Eq.~\ref{eq:mix2_corr}, similar to the STAR measurement. 
The $p$-$k$ mixed cumulant ``true" value can be reproduced via this method, as they are mutually exclusive variables. 
However, the efficiency correction for $Q$-$p$ and $Q$-$k$ fails to reproduce the ``true" values as we discussed in Sec.~\ref{sec:effcorr}. 
We end up with a higher value for unidentified charge correlators. 
This leads to a higher value in cumulant ratios $C_{Q,k} = \sigma_{Q,k}/\sigma^{2}_{k}$ and $C_{Q,p} = \sigma_{Q,p}/\sigma^{2}_{p}$ obtained from ``true" values, as presented in Fig.~\ref{ratioOld}.
This shows that Eq.~\ref{eq:mix2_corr} is not valid for overlap or mutually inclusive variables. 
The observation is qualitatively consistent with the fact that $C_{p,k}$ values in Ref.~\cite{Adam:2019xmk} agree well with the model calculations, while $C_{Q,k}$ and $C_{Q,p}$ are significantly above the model calculations.
However, for mutually exclusive variables (like protons-kaons), there is no issue.
As we discussed before, to correct the mixed cumulant for mutually inclusive variables, Eq.~\ref{eq:mix2_self_correct} is required. 
To apply Eq.~\ref{eq:mix2_self_correct} for $Q$-$k$ and $Q$-$p$ mixed cumulants, it is necessary to identify the charged particles with their efficiencies. 
In Figs.~\ref{offdiaNew} and~\ref{ratioNew}, we solely consider identified charged particles ($Q=\pi+k+p$). 
Then Eq.~\ref{eq:mix2_self_correct} becomes 
\begin{eqnarray}
  \aveave{N_{Q}N_{k}}_{\rm c}
  &=&  \aveave{(N_{\pi}+N_{p}+N_{k})N_{k}}_{\rm c} \nonumber \\ 
  &=& \frac{1}{\varepsilon_{1}\varepsilon_{3}}\ave{N_{\pi} N_{k}} - \frac{1}{\varepsilon_{1}\varepsilon_{3}}\ave{N_{\pi}}\ave{N_{k}} +
  \frac{1}{\varepsilon_{2}\varepsilon_{3}}\ave{N_{p} N_{k}} - \frac{1}{\varepsilon_{2}\varepsilon_{3}}\ave{N_{p}}\ave{N_{k}}
	+ \frac{1}{\varepsilon_{3}^{2}}\ave{N_{k}^{2}} - \frac{1}{\varepsilon_{3}^{2}}\ave{N_{k}}^{2} \nonumber \\ 
	&& + \frac{1}{\varepsilon_{3}}\ave{N_{k}} - \frac{1}{\varepsilon_{3}^{2}}\ave{N_{k}}, 
	\label{eq:mix2_pikp}
\end{eqnarray}
where $\varepsilon_{1}$, $\varepsilon_{2}$, and $\varepsilon_{3}$ are the efficiencies for pions, kaons, and protons, respectively. 
Now, we can reproduce the ``true" values, as presented in Figs.~\ref{offdiaNew} and~\ref{ratioNew}. 

\begin{figure}[htp!]
  \centering
  \subfloat{\includegraphics[width=0.46\textwidth]{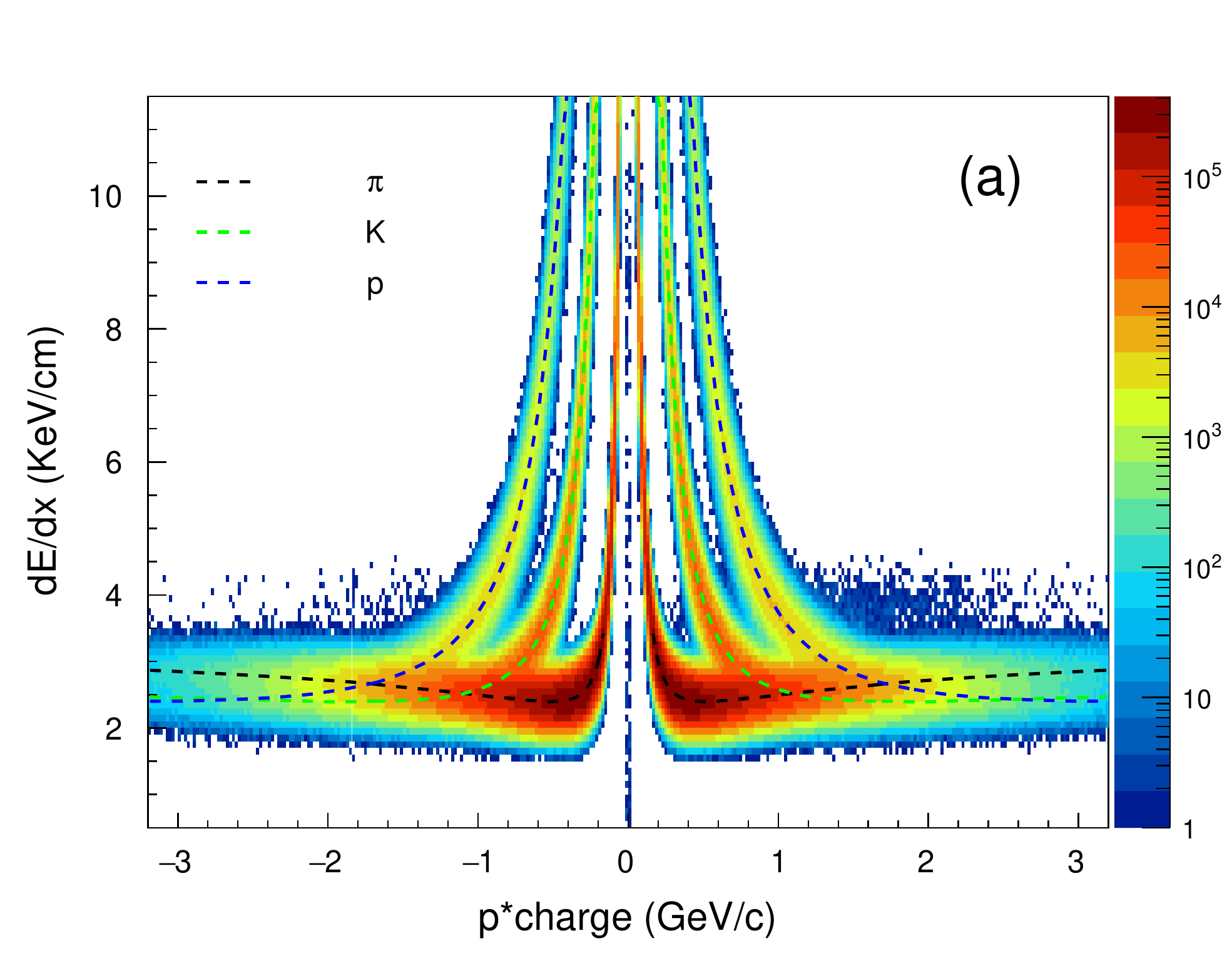}\label{dedx1}}
  \subfloat{\includegraphics[width=0.5\textwidth]{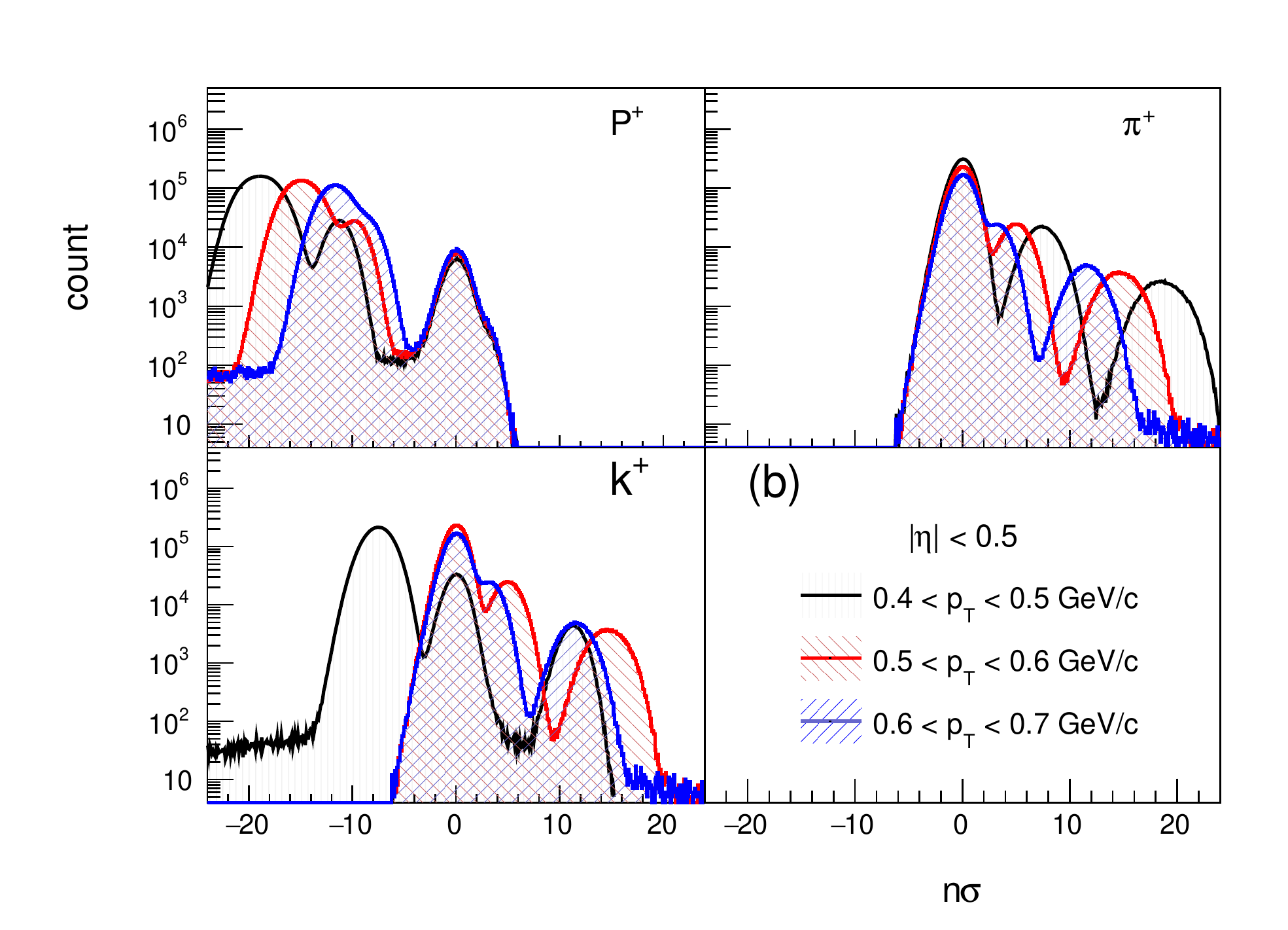}\label{dedx2}}
  \caption{(Color online) (a) $dE/dx$ from UrQMD simulations plotted against charge$\times$momentum of individual particles for Au+Au collisions at \sNN = 200 GeV. (b) $n\sigma$ distributions of protons, pions, and kaons.}
  	\label{dedx}
\end{figure}

However, owing to the particle identification with optimal purity we may lose a few pions, protons, and kaons.  
We also studied the effects of those tracks for mixed cumulants via UrQMD simulations. 
Charged particle identification is performed using the ionization energy loss inside the time projection chamber (TPC) detector subsystem. 
We mimic the ionization energy loss curve in UrQMD simulation using the STAR TPC resolution. 
Figure~\ref{dedx}(a) presents the measured $dE/dx$ distribution after passing the UrQMD input through the TPC simulation. 
The measured values of $dE/dx$ are compared to the expected theoretical values which is an extension of the Bethe-Bloch formula~\cite{Abelev:2008ab} (shown as dashed lines in Fig.~\ref{dedx}(a)). 
To identify particles X, a quantity $n\sigma_{X}$ is defined as 
\begin{eqnarray}
n\sigma_{X} = \frac{1}{R} \ln \frac{[dE/dx]_{obs}}{[dE/dx]_{th,X}},
\end{eqnarray}
where $[dE/dx]_{obs}$ represents the energy loss in the UrQMD simulation and $[dE/dx]_{th,X}$ is the corresponding theoretical value for particle species X. 
R represents the $dE/dx$ resolution, and we use R = 7.5\% within our analysis range. 
Figure~\ref{dedx}(b) presents the $n\sigma$ distributions of $p(\bar{p})$, $\pi^{\pm}$, and $k^{\pm}$. 
Typically, $n\sigma_{p} < 2$ are adopted for the $p(\bar{p})$ selection. Similarly, $n\sigma_{\pi} < 2$ and $n\sigma_{k} < 2$ are adopted for pion and kaon selections, respectively. 
However, from Fig.~\ref{dedx}(a) we can deduce that at high momenta, the $dE/dx$ bands for different particles are overlapped. 
We have also adopted the $2\sigma$-rejection cut to improve the purity.  
\\

\begin{figure*}[htp!]
	\centering 
	\includegraphics[width=0.95\textwidth]{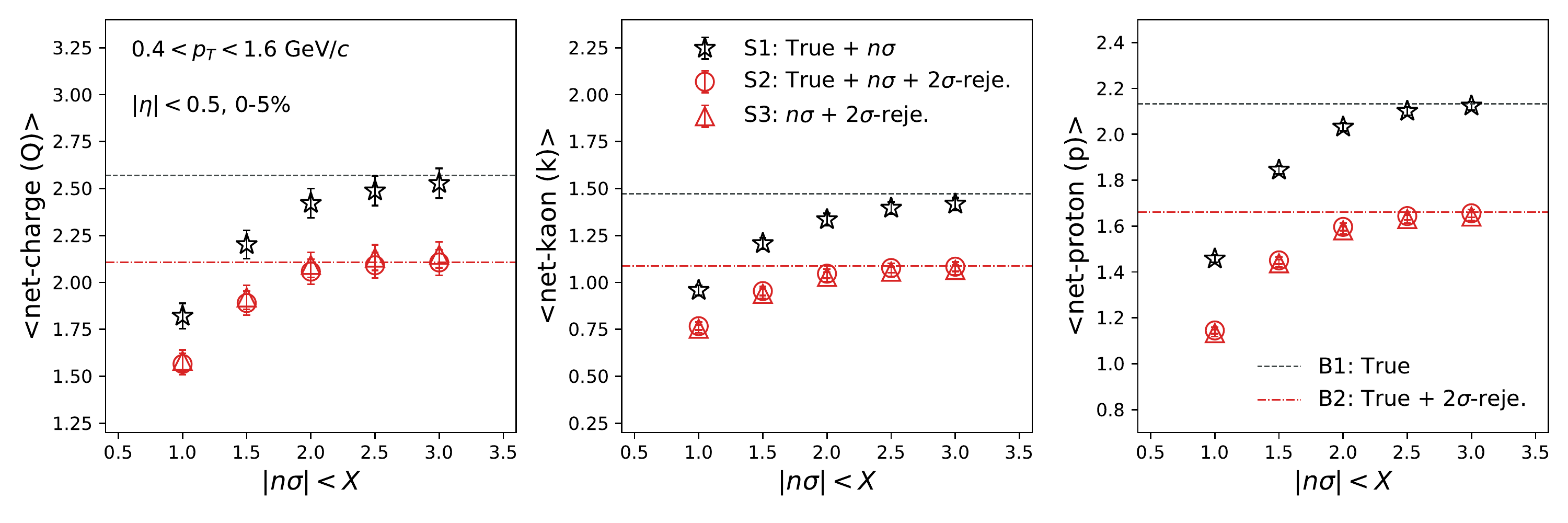}
	\caption{(Color online) $n\sigma$ acceptance dependence mean multiplicity of net-charge (Q), net-proton (p), and net-kaon (k) for Au+Au collisions at 200 GeV, using the UrQMD model.}
	\label{nsigma_mean}
\end{figure*}

\begin{figure*}[htp!]
	\centering 
	\includegraphics[width=0.95\textwidth]{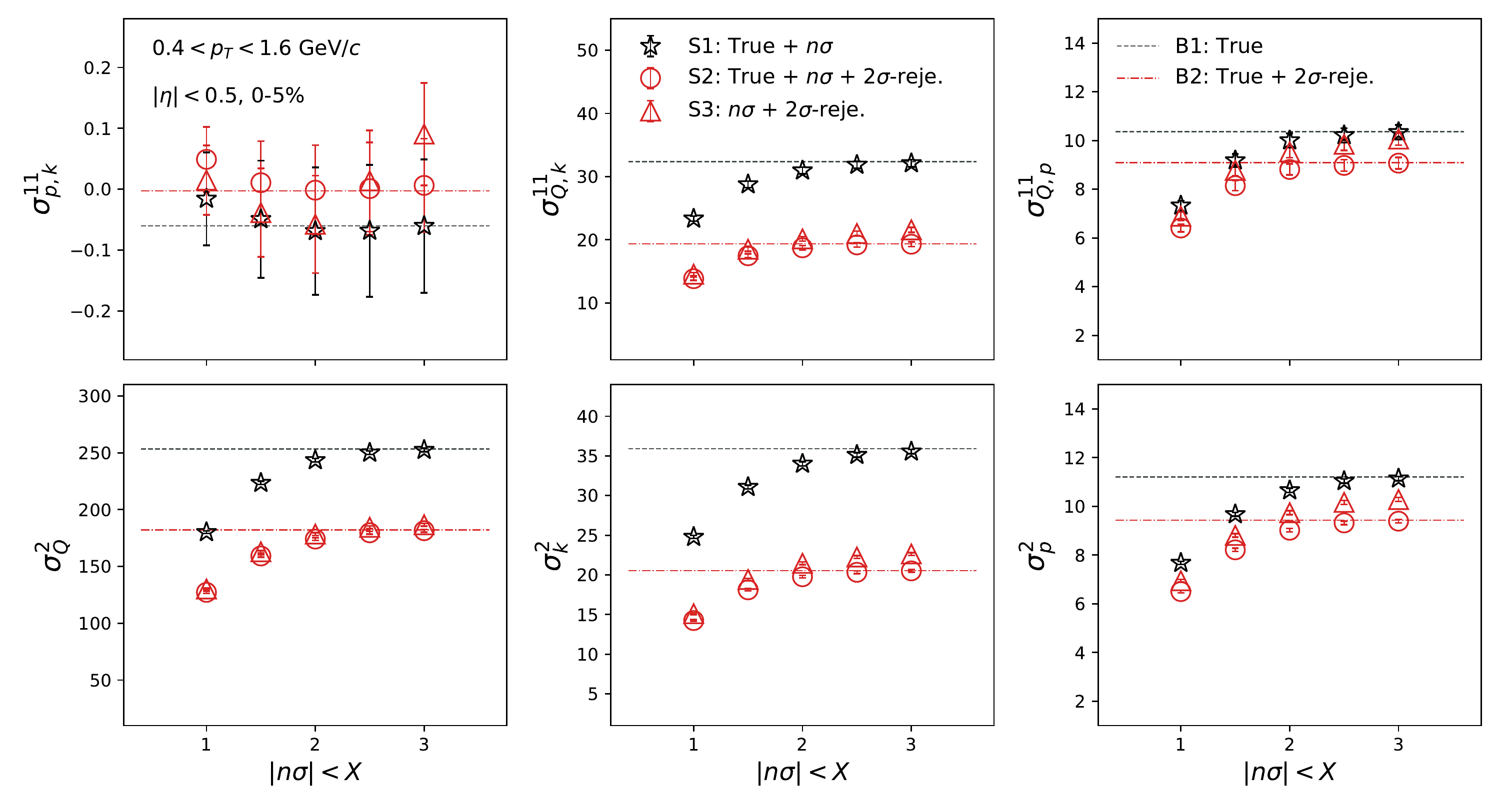}
	\caption{(Color online) $n\sigma$ acceptance dependence second-order off-diagonal and diagonal cumulants for Au+Au collisions at 200 GeV, using the UrQMD model.}
	\label{nsigma_mix}
\end{figure*}


\begin{table}[H]
    \centering
    \begin{tabular}{c|ccccc} \hline
         For & Legend & Particle Code & $n\sigma$ & $2\sigma$-rejection & Baseline  \\ \hline
         \multirow{3}{4em}{Signal} & S1 & Used & Applied & N/A & B1 \\
                & S2 & Used & Applied & Applied & B2 \\
                & S3 & N/A  & Applied & Applied & B2 \\ \hline
        \multirow{2}{4em}{Baseline} & B1 & Used & N/A     & N/A     & N/A \\ 
         & B2 & Used & N/A     & Applied & N/A \\ \hline
    \end{tabular}
    \caption{This table describes the information that is used to select the particles for mixed-cumulant calculations in Fig.~\ref{nsigma_mix}, in terms of the particle species code given by UrQMD, $n\sigma$ and $2\sigma$-rejection cuts. Corresponding legends in Fig.~\ref{nsigma_mean}, ~\ref{nsigma_mix}, and ~\ref{nsigma_mix_norm} are presented in the second left row.}
    \label{tab:tab1}
\end{table}

\begin{figure*}[htp!]
	\centering 
	\includegraphics[width=0.95\textwidth]{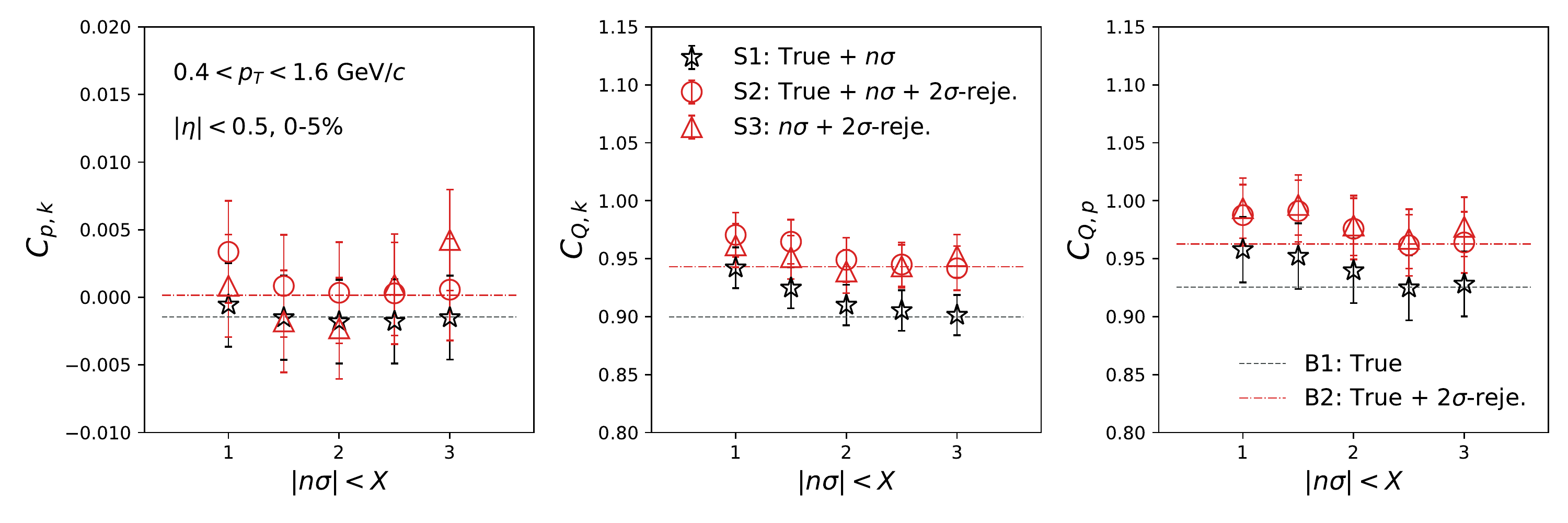}
	\caption{(Color online) $n\sigma$ acceptance dependence second-order off-diagonal over diagonal cumulants ratios for Au+Au collisions at 200 GeV using UrQMD model.}
	\label{nsigma_mix_norm}
\end{figure*}

Figure~\ref{nsigma_mean} presents the event-by-event average of net-charge, net-kaon, and net-proton multiplicities as a function of $n\sigma$. 
  Similarly, figure~\ref{nsigma_mix} shows mixed-cumulants as a function of the $n\sigma$ cut for three cases as follows (refer to Tab.~\ref{tab:tab1}):
  \begin{description}
      \item[S1] : Information on the particle species is provided by UrQMD. The $n\sigma$ cut is applied.
      \item[S2] : Information on the particle species is provided by UrQMD. Both $n\sigma$ and $2\sigma$-rejection cuts are applied. 
      \item[S3] : Particles are identified by both $n\sigma$ and $2\sigma$-rejection cuts. This is the only possible cut in the experimental data analysis. 
  \end{description}
  We note that ``$Q$" is defined as the summation of the identified $\pi$, $K$, and $p$. 
  The ``2$\sigma$-rejection cut" indicates the requirement of $n\sigma_{K} > 2.0$ and $n\sigma_{\pi}>2.0$ for proton identification, $n\sigma_{p}<-2.0$ and $n\sigma_{\pi}>2.0$ for kaon identification, and $n\sigma_{K}<-2.0$ and $n\sigma_{p}<-2.0$ for pion identification. 
  Furthermore, two baselines are calculated with the following conditions, which are independent of the $n\sigma$ cut:
  \begin{description}
      \item[B1] : Information on the particle species is provided by UrQMD.
      \item[B2] : Information on the particle species is provided by UrQMD. The $2\sigma$-rejection cut is applied.
  \end{description}
  The difference between two baselines is owing to the particle multiplicity. There are more particles for B1 than B2 because the $2\sigma$-rejection cut is applied to the latter case.
  Depending on whether the $2\sigma$-rejection cut is applied, 
  S1 can compared to B1, while S2 and S3 need to be compared to B2. 
  The descriptions of Fig.~\ref{nsigma_mix} are summarized in Tab.~\ref{tab:tab1}. 
  
  It is determined that the values of mixed-cumulants are close to the corresponding baselines 
  with a loose $n\sigma$ cut, and decreases with the tightening the $n\sigma$ cut. 
  As can be observed in Eq.~\ref{eq:mix2_true}, the mixed-cumulant for the mutually inclusive case comprises second-order cumulants, as well as second-order mixed-cumulants. 
  It is well known that the former has a trivial volume dependence~\cite{Asakawa:2015ybt}, which can also be confirmed from the lower panels in Fig.~\ref{nsigma_mix} for the second-order cumulants. In Fig.~\ref{nsigma_mix}, therefore, only $\sigma^{11}_{p,k}$ is independent of the $n\sigma$ cut. 
  The difference between S2 and S3 would indicate the effect of contamination. 
  This can be confirmed from the fact that the difference becomes small with the tightening of the $n\sigma$ cut, in addition to the 2$\sigma$-rejection cut. 
  Consequently we observed larger values of $\sigma^{11}_{Q,k}$ and $\sigma^{11}_{Q,p}$ for S3, compared to those for S2.
  
  To cancel the trivial volume dependence, 
  the normalized mixed-cumulants are also calculated in Fig.~\ref{nsigma_mix_norm} as a function of the $n\sigma$ cut. All the results for S1, S2, and S3 are inferred to be consistent with corresponding baselines within uncertainties.  
  This is because the probability of the contamination has been significantly suppressed by the $2\sigma$-rejection cut. To confirm this, normalized mixed-cumulants with only the $n\sigma$ cuts are calculated, where the significant deviations are observed from the corresponding baselines, owing to contamination.
  It should be noted that the $n\sigma$ dependence of mixed-cumulants and normalized mixed-cumulants also depends on how the intrinsic correlations between two variables change relative to the $n\sigma$ cut. 
  In the current simulation, the energy losses of particles are randomly smeared to implement the resolution of the detectors. 
  Therefore, the correlation terms (the first term on the right hand side of Eqs.~\ref{eq:mix2_true} and \ref{eq:mix2_self_true}) are considered to be unaffected by the $n\sigma$ cut. 
  However, experimentally, this effect needs to be carefully studied by changing the criteria for particle identifications.
  
\section{Summary\label{sec:Summary}}
In this study, we discussed the efficiency correction problem for mixed-cumulants. 
This study provided a comprehensive extension of the binomial efficiency correction formula for second-order mixed accumulators in two different cases: one case is for mutually exclusive variables and the other is for mutually inclusive variables. 
We infer that different efficiency correction formulas need to be applied to mixed-cumulants, depending on the type of variable pairs. 
To apply the binomial efficiency correction for $Q$-$k$ and $Q$-$p$ mixed cumulants, it is necessary to identify the charged particles with their corresponding efficiencies. 

It should be noted that the efficiency correction for mixed-cumulants in the case of mutually inclusive variables has already been discussed in Ref.~\cite{Vovchenko:2021xcs}. In the proposed formulas, two different levels of efficiencies 
were implemented for each variable, such as $N_{Q}$ and $N_{p}$ in the case of $\sigma^{11}_{Q,p}$. The tracking efficiency was applied to $N_{Q}$, while the proton identification efficiency was applied on top of the 
tracking efficiency for $N_{p}$.
The method has the advantage that we can keep charged particles as much as possible without identifying each particle species contained in $N_{Q}$. 
This implies that the averaged efficiencies for pions, kaons, and 
protons were used for $N_{Q}$. 
However, we must remember that using the averaged efficiency does not provide the true solution, which depends on underlying probability distributions of the number of particles and the difference in efficiency between particle species~\cite{Nonaka:2017kko}. 
It is also important to note that the identity method would be useful for the 
measurements of mixed-cumulants~\cite{Gazdzicki:2019rrq}, which enables us to measure fluctuations without the multiplicity loss, owing to particle identification. 
It would be desirable for novel ideas to address the two-step efficiency based on the identity method~\cite{Gazdzicki:2019rrq}. 

At this stage, the identification of each particle species and subsequent implemention of the appropriate efficiency is the simplest approach. Therefore, we further investigated the effect of the loss in multiplicity owing to particle identifications, using  numerical simulations.
In the case of mutually inclusive variables, the mixed-cumulants exhibited a monotonic decrease as the cut value of particle identification tightened. This can be explained by a trivial volume dependence.
In contrast, the normalized mixed-cumulants were found to be independent of the cut value for the particle identification. 
This is because the intrinsic correlations between different particle species were assumed to be independent of the variables of the particle identification, which could not be the case in real experiments.
Therefore, it is recommended to verify these effects by changing the criteria for the particle identifications. This work provides an important reference for future measurements of mixed-cumulants in relativistic heavy-ion collisions.

\section{Acknowledgement}
We thank Volker Koch, Masakiyo Kitazawa, Nihar Rajan Sahoo, Prithwish Tribedy, Volodymyr Vovchenko, Tapan Nayak, and Nu Xu for stimulating discussions. This work is supported by the National Key Research and Development Program of China (Grant No. 2020YFE0202002 and 2018YFE0205201), the National Natural Science Foundation of China (Grant No. 11828501, 11890711 and 11861131009), Ito Science Foundation and JSPS KAKENHI Grant No. 25105504, 19H05598.

\bibliography{main}

\appendix
\section{Efficiency correction for higher-order mixed-cumulants}
The efficiency correction formulas for higher-order mixed-cumulants are provided in Ref.~\cite{Nonaka:2017kko} as
\begin{eqnarray}
	  \aveave{ K_{(x)}^2K_{(y)} }_{\rm c}
		&=& \ave{\kappa{(1,0,1)}^{2}\kappa{(0,1,1)}}_{\rm c} 
		+ 2\ave{\kappa{(1,0,1)}\kappa{(1,1,1)}}_{\rm c} 
		- 2\ave{\kappa{(1,0,1)}\kappa{(1,1,2)}}_{\rm c} 
			\nonumber \\ &&
		+ \ave{\kappa{(0,1,1)}\kappa{(2,0,1)}}_{\rm c} 
		- \ave{\kappa{(0,1,1)}\kappa{(2,0,2)}}_{\rm c} 
		+ \ave{\kappa{(2,1,1)}}_{\rm c} -3\ave{\kappa{(2,1,2)}}_{\rm c} + 2\ave{\kappa{(2,1,3)}}_{\rm c},
	  \\ 
	  \nonumber \\ 
	  \aveave{ K_{(x)}^2K_{(y)}^2 }_{\rm c}
		&=& \ave{\kappa{(1,0,1)}^{2}\kappa{(0,1,1)}^{2}}_{\rm c}
		\nonumber \\ &&
		+ \ave{\kappa{(1,0,1)}^{2}\kappa{(0,2,1)}}_{\rm c}
		- \ave{\kappa{(1,0,1)}^{2}\kappa{(0,2,2)}}_{\rm c}
		+ \ave{\kappa{(0,1,1)}^{2}\kappa{(2,0,1)}}_{\rm c}
		- \ave{\kappa{(0,1,1)}^{2}\kappa{(2,0,2)}}_{\rm c}
		\nonumber \\ &&
		+4\ave{\kappa{(1,0,1)}\kappa{(0,1,1)}\kappa{(1,1,1)}}_{\rm c}
		-4\ave{\kappa{(1,0,1)}\kappa{(0,1,1)}\kappa{(1,1,2)}}_{\rm c}
		\nonumber \\ &&
		+2\ave{\kappa{(1,0,1)}\kappa{(1,2,1)}}_{\rm c}
		-6\ave{\kappa{(1,0,1)}\kappa{(1,2,2)}}_{\rm c}
		+4\ave{\kappa{(1,0,1)}\kappa{(1,2,3)}}_{\rm c}
		\nonumber \\ &&
		+2\ave{\kappa{(0,1,1)}\kappa{(2,1,1)}}_{\rm c}
		-6\ave{\kappa{(0,1,1)}\kappa{(2,1,2)}}_{\rm c}
		+4\ave{\kappa{(0,1,1)}\kappa{(2,1,3)}}_{\rm c}
		\nonumber \\ &&
		-4\ave{\kappa{(1,1,1)}\kappa{(1,1,2)}}_{\rm c}
		+2\ave{\kappa{(1,1,1)}^{2}}_{\rm c}
		+2\ave{\kappa{(1,1,2)}^{2}}_{\rm c}
		\nonumber \\ &&
		+ \ave{\kappa{(2,0,1)}\kappa{(0,2,1)}}_{\rm c}
		- \ave{\kappa{(2,0,1)}\kappa{(0,2,2)}}_{\rm c}
		- \ave{\kappa{(2,0,2)}\kappa{(0,2,1)}}_{\rm c}
		+ \ave{\kappa{(2,0,2)}\kappa{(0,2,2)}}_{\rm c}
		\nonumber \\ &&
		+ \ave{\kappa{(2,2,1)}}_{\rm c}
		-7\ave{\kappa{(2,2,2)}}_{\rm c}
		+12\ave{\kappa{(2,2,3)}}_{\rm c}
		-6\ave{\kappa{(2,2,4)}}_{\rm c}, \\ 
	  \nonumber \\ 
	  \aveave{ K_{(x)}^3 K_{(y)} }_{\rm c}
	  &=& \ave{\kappa{(1,0,1)}^{3}\kappa{(0,1,1)}}_{\rm c}
		\nonumber \\ &&
		+3\ave{\kappa{(1,0,1)}^{2}\kappa{(1,1,1)}}_{\rm c}
		-3\ave{\kappa{(1,0,1)}^{2}\kappa{(1,1,2)}}_{\rm c}
		+3\ave{\kappa{(2,0,1)}\kappa{(1,0,1)}\kappa{(0,1,1)}}_{\rm c}
			\nonumber \\ &&
		-3\ave{\kappa{(2,0,2)}\kappa{(1,0,1)}\kappa{(0,1,1)}}_{\rm c}
		+3\ave{\kappa{(1,0,1)}\kappa{(2,1,1)}}_{\rm c}
		-9\ave{\kappa{(1,0,1)}\kappa{(2,1,2)}}_{\rm c}
			\nonumber \\ &&
		+6\ave{\kappa{(1,0,1)}\kappa{(2,1,3)}}_{\rm c}
		+3\ave{\kappa{(2,0,1)}\kappa{(1,1,1)}}_{\rm c}
		-3\ave{\kappa{(2,0,1)}\kappa{(1,1,2)}}_{\rm c}
		-3\ave{\kappa{(2,0,2)}\kappa{(1,1,1)}}_{\rm c}
			\nonumber \\ &&
		+3\ave{\kappa{(2,0,2)}\kappa{(1,1,2)}}_{\rm c}
		+ \ave{\kappa{(3,0,1)}\kappa{(0,1,1)}}_{\rm c}
		-3\ave{\kappa{(3,0,2)}\kappa{(0,1,1)}}_{\rm c}
		+2\ave{\kappa{(3,0,3)}\kappa{(0,1,1)}}_{\rm c}
		\nonumber \\ &&
		+ \ave{\kappa{(3,1,1)}}_{\rm c}
		-7\ave{\kappa{(3,1,2)}}_{\rm c}
		+12\ave{\kappa{(3,1,3)}}_{\rm c}
		-6\ave{\kappa{(3,1,4)}}_{\rm c}.
\end{eqnarray}
The substitution of appropriate indices for $x$ and $y$ in Eq.~\ref{eq:multi_Q} is required, as discussed in Sec.~\ref{sec:effcorr}.

\end{document}